\documentclass[11pt]{article}

\usepackage[utf8]{inputenc}
\usepackage{cite}  
\usepackage{graphicx}
\usepackage[margin=1.25in]{geometry}
\usepackage[usenames,dvipsnames]{color}
\usepackage{url}
\usepackage[colorlinks = true,
            linkcolor = blue,
            urlcolor  = blue,
            citecolor = blue,
            anchorcolor = blue]{hyperref}
            
\usepackage{orcidlink}
\usepackage{lineno}
\usepackage{wrapfig}
\usepackage{float}
\usepackage[caption=false]{subfig}

\def\Title#1{\begin{center} {\LARGE #1 } \end{center}}

\newenvironment{Abstract}{\begin{quotation} \begin{center}
                       ABSTRACT
     \end{center}\bigskip  }{\end{quotation}}

\newcommand\snowmass{\begin{center}\rule[-0.2in]{\hsize}{0.01in}\\\rule{\hsize}{0.01in}\\
\vskip 0.1in Submitted to the  Proceedings of the US Community Study\\ 
on the Future of Particle Physics (Snowmass 2021)\\ 
\rule{\hsize}{0.01in}\\\rule[+0.2in]{\hsize}{0.01in} \end{center}}

\def\beq{\begin{equation}}
\def\eeq#1{\label{#1}\end{equation}}
\def\eeqn{\end{equation}}

\newenvironment{Eqnarray}%
   {\arraycolsep 0.14em\begin{eqnarray}}{\end{eqnarray}}
\def\beqa{\begin{Eqnarray}}
\def\eeqa#1{\label{#1}\end{Eqnarray}}
\def\eeqan{\end{Eqnarray}}

\let\bar=\overbar

\def\lsim{\mathrel{\raise.3ex\hbox{$<$\kern-.75em\lower1ex\hbox{$\sim$}}}}
\def\gsim{\mathrel{\raise.3ex\hbox{$>$\kern-.75em\lower1ex\hbox{$\sim$}}}}

\def\del{\partial}
\def\Dslash{\not{\hbox{\kern-4pt $D$}}}
\def\dslash{\not{\hbox{\kern-2pt $\del$}}}
\def\pslash{\not{\hbox{\kern-2pt $p$}}}
\def\ETmiss{\not{\hbox{\kern-4pt $E$}}_T}

\def\Dlr{\mathrel{\raise1.5ex\hbox{$\leftrightarrow$\kern-1em\lower1.5ex\hbox{$D$}}}}

\def\MSB{{\bar{M \kern -2pt S}}}
\def\msb{{\bar{\scriptsize M \kern -1pt S}}}

\def\drb{{\bar{\scriptsize D \kern -1pt R}}}

\begin{document}


\snowmass{}

\Title{\textbf{Explainable AI for High Energy Physics}\\
\vspace{10pt}%
COMPF3 (Machine Learning)}

\begin{center} 

Mark S. Neubauer\,\orcidlink{0000-0001-8434-9274}\textsuperscript{1},
Avik Roy\,\orcidlink{0000-0002-0116-1012}\textsuperscript{1}
\end{center}

\vspace{-18pt}
\begin{center}
\textbf{1}~University of Illinois at Urbana-Champaign, Urbana IL 61801, USA
\end{center}

\begin{Abstract}
\vspace{-4pt}
\noindent
Neural Networks are ubiquitous in high energy physics research. However, these highly nonlinear parameterized functions are treated as \textit{black boxes}- whose inner workings to convey information and build the desired input-output relationship are often intractable. Explainable AI (xAI) methods can be useful in determining a neural model's relationship with data toward making it \textit{interpretable} by establishing a quantitative and tractable relationship between the input and the model's output. In this letter of interest, we explore the potential of using xAI methods in the context of problems in high energy physics. 
\end{Abstract}

\pagebreak

\section{Introduction}

The evolution of machine learning machine learning models in high energy physics (HEP) research can be attributed to three factors- (i) growing intricacy in detector design and operation, (ii) big data of unprecedented volumes, and (iii) availability of fast computing resources.  Over the years, simpler and interpretable regression and classification models have been replaced by intractable, \textit{black-box}-like deep neural networks. 
Recent progress in the field of \textit{explainabale} Artificial Intelligence (xAI)~\cite{MILLER20191} have made it possible to investigate the relationships between an AI model's inputs, architecture, and predictions~\cite{xAI-intro,xAI-review,xAI-review-2}. Some methods remain model agnostic, while other methods have been developed to interpret  computer vision models where intuitive reasoning can be extracted from human-annotated datasets to validate xAI techniques. However, in other data structures, like large tabular data or relational data constructs like graphs, use of xAI methods are still quite novel~\cite{sahakyan2021explainable, xAI-GNN-survey}. The scope of xAI in high energy physics~\cite{9302535} has been rather limited but not without success- showing promise, for example, in learning parton showers at the LHC~\cite{LAI2022137055} and jet reconstruction using particle flow algorithms~\cite{mokhtar2021explaining}.

\section{The Scope of xAI in High Energy Physics}
\label{sec:scope}

Explainability may not be indispensable for many ML tasks. However, in most such cases, either the model itself is simple enough that explanations are rather trivial or the impact of such models remains marginal in a broader problem setting. If we are to understand, trust, and manage AI models, it is important that a humanly intelligible interpretation should exist~\cite{MILLER20191}. Experimental high energy physics deals with incredibly complex detectors and very large datasets. Neural networks and deep learning models already play very important roles in analysis of detector simulation and performance, reconstruction of physics objects from tracks and energy deposits in the detectors~\cite{ciodaro2012online, pata2021mlpf}, modeling of parton distribution and showers~\cite{carrazza2019towards}, and dedicated jet tagging to identify boosted, heavy quarks~\cite{aad2019atlasbjet}. Given the large computational expense of traditional detector simulation algorithms, generative models for accelerated simulation have been explored~\cite{salamani2018deep}. The particle physics community has also started exploring the implementation of pre-trained DNN models on FPGA devices for online and offline trigger systems~\cite{duarte2018fast}, object reconstruction~\cite{iiyama2021distance}, tracking and tagging~\cite{heintz2020accelerated, thais2022graph}. The spectrum of ML applications in LHC research is broad and ever-increasing. Given their intimate usage with online detector operations, it becomes increasingly more important to be able to explain and interpret these models.

    \subsection{Generative models for physics and detector simulation}
        Using generative models for detector simulation has shown initial success in speeding up the process of simulation with commendable accuracy~\cite{salamani2018deep,vallecorsa2018generative, hashemi2019lhc}. However, generative models are known to have some limitations. For instance, Generative Adversarial Networks (GANs) are known to cause mode collapse where a large subspace of latent space embedding maps to a relatively small subset of the feature space~\cite{bau2019seeing}. On the other hand, Variational Auto Encoders (VAEs) often lose sharpness in the generated model space~\cite{dai2019diagnosing}. Application of such models for LHC models hence requires rigorous scrutiny since physics of interest are rare and their latent space embeddings need to be well understood to make sure physics details and their resolutions are not lost in the process. How the latent space embeddings rely on input features, how well the network recognizes and manages correlated input features- these questions need to be understood in the context of such machine learning models.
        
    \subsection{Machine learning for object reconstruction}
        Combining millions of detector signals to reconstruct objects of interest, such as electrons, muons, and jets relies on dedicated algorithms for track identification and track matching with calorimeter signatures. Recent developments have shown that ML models can successfully recreate and sometimes outperform traditional algorithms~\cite{pata2021mlpf}. How these models evaluate the input data and weigh the relative importance of these features need to be understood. For instance, the feature importance for MLPFlow algorithm has been explored by using the Layerwise Relevance Propagation (LRP) algorithm~\cite{mokhtar2021explaining}.
    
    \subsection{Modeling of parton distribution and showers}
        Effective modeling the partonic structure of colliding hadrons, as done by the \texttt{NNPDF} collaboration for instance, has successfully employed DNN techniques~\cite{duarte2018fast,carrazza2019towards}. The underlying training code and methods have been recently made public~\cite{ball2021open}. DNNs have also been used to investigate parton showers dictated by non-perturbative QCD~\cite{bothmann2019reweighting}. Recent results with xAI methods have shown that such models are capable of not only  correctly predicting the final distribution of particles but also learning the underlying physics~\cite{LAI2022137055}. Extensive efforts in understanding ML models' capability of learning QCD physics to correctly understand physics models for HL-LHC and future detectors can significantly improve discovery potential and analysis sensitivity.
    
    \subsection{Object tagging and event classification}
        This is one of the most active areas of ML application in the experimental HEP. Identifying potential signal events from a much larger set of background events- essentially a needle in a haystack problem- makes it ideal to adapt already tested and validated models from computer vision and natural language processing. For instance, the \texttt{GoogleNET} model~\cite{szegedy2015going}, a convolutional neural network (CNN) for visual image recognition, was successfully adapted for classifying neutrinoless double beta decay with the NEXT-100~\cite{alvarez2013near} detector~\cite{renner2017background}. In the context of the NOvA detector~\cite{ayres2007nova}, similar studies were done with modified CNN models~\cite{aurisano2016convolutional}. While such applications of established ML models in neutrino experiments have made them increasingly more popular over the years, there are major challenges, including interpretability of these models, associated with their comprehensibility and trustworthiness~\cite{psihas2020review}. 
        
\begin{figure}[!h]
\centering
\subfloat[]{
\includegraphics[width=0.5\textwidth]{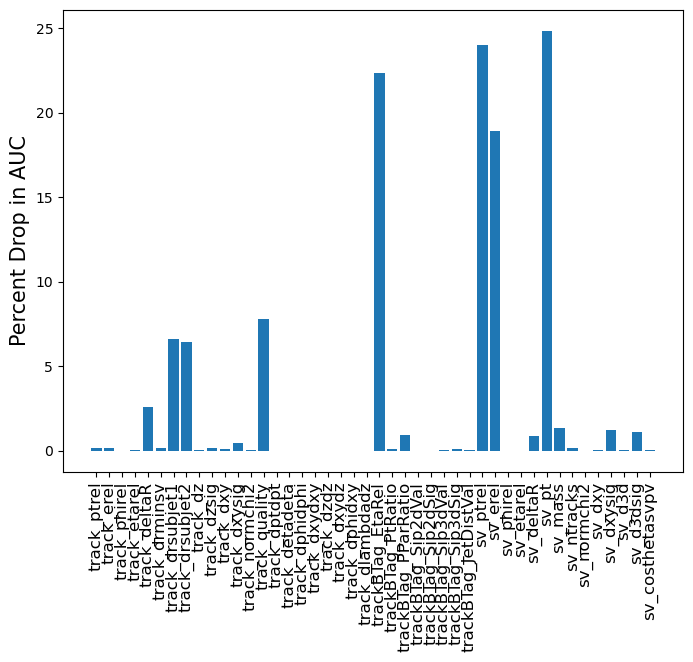}
\label{fig:dAUC-baseline}            
}
\subfloat[]{
\includegraphics[width=0.5\textwidth]{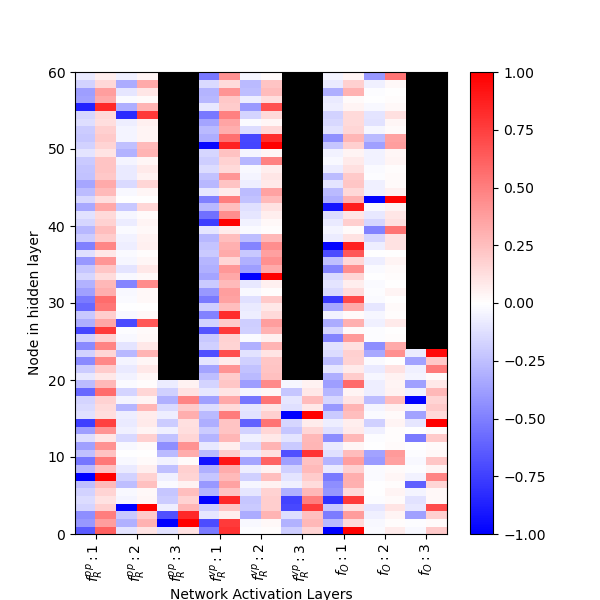}
\label{fig:NAP-baseline}            
}
\caption{\protect\subref{fig:dAUC-baseline} Change in AUC score with respect to the baseline model when each of the  track and  secondary vertex features is individually masked during inference with the trained baseline model.
\protect\subref{fig:NAP-baseline} 2D map of relative neural activity for different nodes of the activation layers. To simultaneously visualize the scores for QCD and $H \rightarrow b\bar{b}$ jets, we project the activity scores of the former as negative values.}
\label{fig:IN-xAI}
\end{figure}

        On the energy frontier, identifying the origin of jets and associating them with partons is one of the areas where ML has been most widely used. Ref.~\cite{kasieczka2019machine} summarizes a wide variety of models dedicated to idenfity jets originating from top quarks, novel architectures like Interaction Network (IN)~\cite{IN} have been developed to identify $H\rightarrow b\bar{b}$ jets from QCD background. The inputs to these models are a large set of kinematic variables associated with particle tracks, secondary vertices, and calorimeter signatures. Incorporating such large number of variables that have non-trivial, intractable correlations can lead to arbitrarily complex models while models themselves can rely on a handful of these features. For instance, we examine the change in ROC-AUC score~\cite{huang2005using} for the IN model when individual particle track and secondary vertex features are masked in Figure~\ref{fig:dAUC-baseline}. It reveals that the model's performance is hardly affected when certain features are masked. Figure~\ref{fig:NAP-baseline} shows the Neural Activation Pattern (NAP) diagram for the model, showing relative activation strength of each node in the activation layers of the model, each normalized with respect to the largest cumulative activation node in the same layer. The relative activation strength illustrates that only a smaller subset of the activation nodes strongly participate in information propagation across the model. These studies suggest that the model can be made significantly simpler, both in terms on the number of features is relies on as well as the number of trainable parameters.
        
\subsection{DNNs built on FPGAs for Trigger applications}
        With DNN's success in jet and event classification, recent work has placed emphasis on developing DNN-enabled FPGAs for trigger level applications at the LHC~\cite{duarte2018fast,iiyama2021distance,heintz2020accelerated,thais2022graph}. Since these inference-only dedicated hardware can play crucial role in event classification and identifying events of interest in real time, explainability becomes an important criteria to ensure these models are trustworthy. Interpretability of DNNs implemented on decdicated hardware requires simultaneous understanding of how a model's performance varies when operating on these platforms as well as how the model's architecture interplays with the electonic signals received from the detectors.

\section{Summary and Outlook}
The scope of xAI in HEP research is as broad as that of ML applications in HEP. While the existing methods of xAI can help us better understand, trust, and optimize the various ML models used in HEP research, the constraints from underlying physics may also pave ways of developing novel explainability metrics for dedicated ML models in HEP. Even beyond the usual scope of regression and classification tasks, such dedicated xAI methods can be very useful for uncertainty quantification. The interface of xAI and uncertainty quantification in ML is still quite unexplored, both as a broader statistical problem and in domain specific applications~\cite{seuss2021bridging}. The broader scope of ML applications in HEP will require dedicated understanding of how robust as well as interpretable these models are~\cite{grojean2022lessons}. Being a data intensive research community, methods of xAI will provide invaluable insight regarding how data relates with intricate and otherwise intractable models that represent them. Hence, it is quite fathomable that the community's major scientific drives, both in near- and long-term, will significantly benefit from dedicated exploration of such methods.  

\bibliography{main}

\begin{thebibliography}{10}

\bibitem{MILLER20191}
Tim Miller.
\newblock Explanation in artificial intelligence: Insights from the social
  sciences.
\newblock {\em Artificial Intelligence}, 267:1--38, 2019.

\bibitem{xAI-intro}
David Gunning, Mark Stefik, Jaesik Choi, Timothy Miller, Simone Stumpf, and
  Guang-Zhong Yang.
\newblock Xai—explainable artificial intelligence.
\newblock {\em Science Robotics}, 4(37):eaay7120, 2019.

\bibitem{xAI-review}
Pantelis Linardatos, Vasilis Papastefanopoulos, and Sotiris Kotsiantis.
\newblock Explainable ai: a review of machine learning interpretability
  methods.
\newblock {\em Entropy}, 23(1):18, 2020.

\bibitem{xAI-review-2}
Giulia Vilone and Luca Longo.
\newblock Explainable artificial intelligence: a systematic review.
\newblock {\em arXiv preprint arXiv:2006.00093}, 2020.

\bibitem{sahakyan2021explainable}
Maria Sahakyan, Zeyar Aung, and Talal Rahwan.
\newblock Explainable artificial intelligence for tabular data: A survey.
\newblock {\em IEEE Access}, 9:135392--135422, 2021.

\bibitem{xAI-GNN-survey}
Hao Yuan, Haiyang Yu, Shurui Gui, and Shuiwang Ji.
\newblock Explainability in graph neural networks: A taxonomic survey.
\newblock {\em arXiv preprint arXiv:2012.15445}, 2020.

\bibitem{9302535}
Danielle Turvill, Lee Barnby, Bo~Yuan, and Ali Zahir.
\newblock A survey of interpretability of machine learning in accelerator-based
  high energy physics.
\newblock In {\em 2020 IEEE/ACM International Conference on Big Data Computing,
  Applications and Technologies (BDCAT)}, pages 77--86, 2020.

\bibitem{LAI2022137055}
Yue~Shi Lai, Duff Neill, Mateusz Płoskoń, and Felix Ringer.
\newblock Explainable machine learning of the underlying physics of high-energy
  particle collisions.
\newblock {\em Physics Letters B}, 829:137055, 2022.

\bibitem{mokhtar2021explaining}
Farouk Mokhtar, Raghav Kansal, Daniel Diaz, Javier Duarte, Joosep Pata,
  Maurizio Pierini, and Jean-Roch Vlimant.
\newblock Explaining machine-learned particle-flow reconstruction.
\newblock {\em arXiv preprint arXiv:2111.12840}, 2021.

\bibitem{ciodaro2012online}
T~Ciodaro, D~Deva, JM~De~Seixas, and D~Damazio.
\newblock Online particle detection with neural networks based on topological
  calorimetry information.
\newblock In {\em Journal of physics: conference series}, volume 368, page
  012030. IOP Publishing, 2012.

\bibitem{pata2021mlpf}
Joosep Pata, Javier Duarte, Jean-Roch Vlimant, Maurizio Pierini, and Maria
  Spiropulu.
\newblock Mlpf: efficient machine-learned particle-flow reconstruction using
  graph neural networks.
\newblock {\em The European Physical Journal C}, 81(5):1--14, 2021.

\bibitem{carrazza2019towards}
Stefano Carrazza and Juan Cruz-Martinez.
\newblock Towards a new generation of parton densities with deep learning
  models.
\newblock {\em The European Physical Journal C}, 79(8):1--9, 2019.

\bibitem{aad2019atlasbjet}
{The ATLAS Collaboration}.
\newblock Atlas b-jet identification performance and efficiency measurement
  with $t\bar{t}$ events in pp collisions at $\sqrt{s}= 13$ tev.
\newblock {\em The European physical journal C}, 79(11):1--36, 2019.

\bibitem{salamani2018deep}
Dalila Salamani, Stefan Gadatsch, Tobias Golling, Graeme~Andrew Stewart, Aishik
  Ghosh, David Rousseau, Ahmed Hasib, and Jana Schaarschmidt.
\newblock Deep generative models for fast shower simulation in atlas.
\newblock In {\em 2018 IEEE 14th International Conference on e-Science
  (e-Science)}, pages 348--348. IEEE, 2018.

\bibitem{duarte2018fast}
Javier Duarte, Song Han, Philip Harris, Sergo Jindariani, Edward Kreinar,
  Benjamin Kreis, Jennifer Ngadiuba, Maurizio Pierini, Ryan Rivera, Nhan Tran,
  et~al.
\newblock Fast inference of deep neural networks in fpgas for particle physics.
\newblock {\em Journal of Instrumentation}, 13(07):P07027, 2018.

\bibitem{iiyama2021distance}
Yutaro Iiyama, Gianluca Cerminara, Abhijay Gupta, Jan Kieseler, Vladimir
  Loncar, Maurizio Pierini, Shah~Rukh Qasim, Marcel Rieger, Sioni Summers,
  Gerrit Van~Onsem, et~al.
\newblock Distance-weighted graph neural networks on fpgas for real-time
  particle reconstruction in high energy physics.
\newblock {\em Frontiers in big Data}, page~44, 2021.

\bibitem{heintz2020accelerated}
Aneesh Heintz, Vesal Razavimaleki, Javier Duarte, Gage DeZoort, Isobel Ojalvo,
  Savannah Thais, Markus Atkinson, Mark Neubauer, Lindsey Gray, Sergo
  Jindariani, et~al.
\newblock Accelerated charged particle tracking with graph neural networks on
  fpgas.
\newblock {\em arXiv preprint arXiv:2012.01563}, 2020.

\bibitem{thais2022graph}
Savannah Thais, Paolo Calafiura, Grigorios Chachamis, Gage DeZoort, Javier
  Duarte, Sanmay Ganguly, Michael Kagan, Daniel Murnane, Mark~S Neubauer, and
  Kazuhiro Terao.
\newblock Graph neural networks in particle physics: Implementations,
  innovations, and challenges.
\newblock {\em arXiv preprint arXiv:2203.12852}, 2022.

\bibitem{vallecorsa2018generative}
Sofia Vallecorsa.
\newblock Generative models for fast simulation.
\newblock In {\em Journal of Physics: Conference Series}, volume 1085, page
  022005. IOP Publishing, 2018.

\bibitem{hashemi2019lhc}
Bobak Hashemi, Nick Amin, Kaustuv Datta, Dominick Olivito, and Maurizio
  Pierini.
\newblock Lhc analysis-specific datasets with generative adversarial networks.
\newblock {\em arXiv preprint arXiv:1901.05282}, 2019.

\bibitem{bau2019seeing}
David Bau, Jun-Yan Zhu, Jonas Wulff, William Peebles, Hendrik Strobelt, Bolei
  Zhou, and Antonio Torralba.
\newblock Seeing what a gan cannot generate.
\newblock In {\em Proceedings of the IEEE/CVF International Conference on
  Computer Vision}, pages 4502--4511, 2019.

\bibitem{dai2019diagnosing}
Bin Dai and David Wipf.
\newblock Diagnosing and enhancing vae models.
\newblock {\em arXiv preprint arXiv:1903.05789}, 2019.

\bibitem{ball2021open}
Richard~D Ball, Stefano Carrazza, Juan Cruz-Martinez, Luigi Del~Debbio, Stefano
  Forte, Tommaso Giani, Shayan Iranipour, Zahari Kassabov, Jose~I Latorre,
  Emanuele~R Nocera, et~al.
\newblock An open-source machine learning framework for global analyses of
  parton distributions.
\newblock {\em The European Physical Journal C}, 81(10):1--12, 2021.

\bibitem{bothmann2019reweighting}
Enrico Bothmann and Luigi Del~Debbio.
\newblock Reweighting a parton shower using a neural network: the final-state
  case.
\newblock {\em Journal of High Energy Physics}, 2019(1):1--24, 2019.

\bibitem{szegedy2015going}
Christian Szegedy, Wei Liu, Yangqing Jia, Pierre Sermanet, Scott Reed, Dragomir
  Anguelov, Dumitru Erhan, Vincent Vanhoucke, and Andrew Rabinovich.
\newblock Going deeper with convolutions.
\newblock In {\em Proceedings of the IEEE conference on computer vision and
  pattern recognition}, pages 1--9, 2015.

\bibitem{alvarez2013near}
V~{\'A}lvarez, FIGM Borges, S~C{\'a}rcel, J~Castel, S~Cebri{\'a}n, A~Cervera,
  Carlos~AN Conde, Theopisti Dafni, THVT Dias, J~D{\'\i}az, et~al.
\newblock Near-intrinsic energy resolution for 30--662 kev gamma rays in a high
  pressure xenon electroluminescent tpc.
\newblock {\em Nuclear Instruments and Methods in Physics Research Section A:
  Accelerators, Spectrometers, Detectors and Associated Equipment},
  708:101--114, 2013.

\bibitem{renner2017background}
Joshua Renner, A~Farbin, J~Mu{\~n}oz Vidal, JM~Benlloch-Rodr{\'\i}guez,
  A~Botas, Paola Ferrario, Juan~Jos{\'e} G{\'o}mez-Cadenas, Vicente Alvarez,
  CDR Azevedo, FIG Borges, et~al.
\newblock Background rejection in next using deep neural networks.
\newblock {\em Journal of Instrumentation}, 12(01):T01004, 2017.

\bibitem{ayres2007nova}
DS~Ayres, GR~Drake, MC~Goodman, JJ~Grudzinski, VJ~Guarino, RL~Talaga, A~Zhao,
  P~Stamoulis, E~Stiliaris, G~Tzanakos, et~al.
\newblock The nova technical design report.
\newblock Technical report, Fermi National Accelerator Lab.(FNAL), Batavia, IL
  (United States), 2007.

\bibitem{aurisano2016convolutional}
Adam Aurisano, Alexander Radovic, D~Rocco, Alexander Himmel, MD~Messier,
  E~Niner, G~Pawloski, Fernanda Psihas, Alexandre Sousa, and P~Vahle.
\newblock A convolutional neural network neutrino event classifier.
\newblock {\em Journal of Instrumentation}, 11(09):P09001, 2016.

\bibitem{psihas2020review}
Fernanda Psihas, Micah Groh, Christopher Tunnell, and Karl Warburton.
\newblock A review on machine learning for neutrino experiments.
\newblock {\em International Journal of Modern Physics A}, 35(33):2043005,
  2020.

\bibitem{kasieczka2019machine}
Gregor Kasieczka, Tilman Plehn, Anja Butter, Kyle Cranmer, Dipsikha Debnath,
  Barry~M Dillon, Malcolm Fairbairn, Darius~A Faroughy, Wojtek Fedorko,
  Christophe Gay, et~al.
\newblock The machine learning landscape of top taggers.
\newblock {\em SciPost Physics Proceedings}, 7(1), 2019.

\bibitem{IN}
Eric~A Moreno, Thong~Q Nguyen, Jean-Roch Vlimant, Olmo Cerri, Harvey~B Newman,
  Avikar Periwal, Maria Spiropulu, Javier~M Duarte, and Maurizio Pierini.
\newblock Interaction networks for the identification of boosted h→ b b
  decays.
\newblock {\em Physical Review D}, 102(1):012010, 2020.

\bibitem{huang2005using}
Jin Huang and Charles~X Ling.
\newblock Using auc and accuracy in evaluating learning algorithms.
\newblock {\em IEEE Transactions on knowledge and Data Engineering},
  17(3):299--310, 2005.

\bibitem{seuss2021bridging}
Dominik Seu{\ss}.
\newblock Bridging the gap between explainable ai and uncertainty
  quantification to enhance trustability.
\newblock {\em arXiv preprint arXiv:2105.11828}, 2021.

\bibitem{grojean2022lessons}
Christophe Grojean, Ayan Paul, Zhuoni Qian, and Inga Str{\"u}mke.
\newblock Lessons on interpretable machine learning from particle physics.
\newblock {\em Nature Reviews Physics}, pages 1--3, 2022.

\end{thebibliography}
\bibliographystyle{unsrt}
\end{document}